\input phyzzx
\PHYSREV


\def\a {\alpha}
\def\b {\beta}
\def\g {\gamma}

\def\d {\delta}
\def\e {\epsilon}
\def\f {\psi}

\def\l {\lambda}

\def\s {\sigma}
\def\t {\tau}

\def\vf {\chi}


\def\dif {\rm d}

\def\half {\coeff{1}{2}}

\def\db {\dot {\b}}

\def\x {\xi}
\def\y {\eta}
\def\pde #1#2{\partial_{#2} #1}
\def\z {\phantom z}
\def\s{\sigma}

\REF\BF{V. Belinskii and M. Francaviglia, Gen. Rel. and Grav. 14, 213 (1982) }

\REF\BZ{Belinskii and Zakharov, Sov. Phys. JETP 48, 985 (1978)}




\REF\MM{N. Manojlovi\'c and A. Mikovi\'c, Painleve III Equation and Bianchi 
VII Model, preprint gr-qc/9908077}

\REF\IN{E.L. Ince, Ordinary Differential Equations, Dover Publications, 
New York (1956)}

\REF\JAN{R.T. Janzen, Commun. Math. Phys. 64, 211 (1979) }

\REF\K{H. Kodama, Prog. Theor. Phys. 99, 173 (1998)}

\REF\FN{H. Flaschka and A.C. Newell, Commun. Math. Phys. 76, 65 (1980)}

\REF\JMU{M. Jimbo, T. Miwa and K. Ueno, Physica D 2, 306 (1981)}

\REF\I{A.R. Its and V.Y. Novokshenov, The Isomonodromic Deformation Method in
the Theory of Painlev\'e Equations, Lecture Notes in Mathematics, 
Springer-Verlag, Berlin (1986)}

\REF\MS{N. Manojlovi\'c and B. Spence, Nucl. Phys. B 423, 243 (1994)}

\REF\KS{D. Korotkin and H. Semtleben, Phys. Rev. Lett. 80, 14 (1998) }

\REF\HR{J. Harnad and M. Routhier, J. Math. Phys. 36, 4863 (1995)}

\date={February 2000}

\titlepage

\title{\bf Belinskii-Zakharov Formulation for Bianchi Models and
Painlev\'e III Equation}

\author{Nenad Manojlovi\'c\foot{E-mail:nmanoj@ualg.pt} and 
Aleksandar Mikovi\'c\foot{On leave 
of absence from Institute of Physics, Belgrade, Yugoslavia }
\foot{E-mail:amikovic@ualg.pt}}
\address{\'Area Departamental de Matem\'atica, UCEH, Universidade do Algarve, Campus de 
Gambelas, 8000 Faro, Portugal}

\abstract {We show that $\a<0,\b>0,\g=\d=0$  Painlev\'e III equation 
arises as a zero-curvature condition
in the Belinskii-Zakharov inverse scattering formulation for Bianchi 
cosmological models. For special 
values of the parameters this Painlev\'e III equation becomes the dynamical 
equation for Bianchi I, II, VI$_0$ and VII$_0$ models.}
\endpage

{\bf \chapter {Introduction}}

Belinskii and Francaviglia showed in [\BF] that 
the Einstein equations for Bianchi I, II, VI$_0$ and VII$_0$ spacetimes
admit a zero-curvature 
representation, i.e. they found a linear system whose integrability condition
is the dynamical equation for the Bianchi model. 
This was done by using a more general framework of 
Belinskii-Zakharov (BZ) inverse scattering method  
for the spacetimes admitting two commuting spacelike Killing vectors [\BZ]. 
The results of [\BF]
demonstrated that the Bianchi models which admit two commuting spacelike 
Killing vectors
are solvable dynamical systems. 
However, not much work has been done on the issue of what kind of integrable
nonlinear dynamical equations can be obtained from this approach. 

In [\MM] it has been shown that in the case of the Bianchi VII$_0$ model one
obtains a special Painlev\'e III (PIII) equation, which is 
$\a=-2$, $\b = 2$, $\g =\d =0$ case of the standard PIII form [\IN]
$$
\eqalignno { {d^2 u\over dt^2} &= {1\over u} \Bigl ( {d u\over dt} 
\Bigr )^2 - {1\over t} \Bigl ( {d u\over d t} \Bigr ) + {1\over t}
 \Bigl ( \a u^2 + \b \Bigr )+ \g u^3 + \d u^{-1} \, . &\eqnalign\spiii \cr}
$$
In this paper we consider the Belinskii-Zakharov formulation for all Bianchi 
models which admit two commuting
Killing vectors, i.e. types I, II, VI$_0$ and VII$_0$. We only consider the
equations of motion for local degrees of freedom, and we do not discuss the 
problems related with non-trivial topology of the spatial manifold [\MM].   
By considering a larger class of Bianchi models, we obtain a more general 
PIII equation as the dynamical equation,
namely $\a <0$, $\b >0 $, $\g=\d=0$. Although the relevant Bianchi models
correspond to special values of the parameters, we show that the zero curvature
representation is valid for all other values of the parameters. 
Consequently we obtain a zero-curvature
representation of $\a <0$, $\b > 0$, $\g =\d =0$ Painlev\'e III equation 
in the Belinskii-Zakharov inverse scattering formulation. 

{\bf \chapter {Zero-curvature formulation for Bianchi models}}

Belinskii-Francaviglia approach to solving the dynamics of Bianchi 
models [\BF] is derived from the Belinskii-Zakharov method for
solving the Einstein equations for spacetimes with two commuting spacelike 
Killing vectors [\BZ]. Such spacetimes have the following form of the metric 
$$
\eqalignno{ ds ^2 &=  f(t,z)(-\ dt ^2 +  \ dz ^2) + 
g_{ab}(t,z) \ dx ^a dx ^b \ ,
&\eqnalign\dvakv	\cr}
$$
where $a,b = 1,2$, $\{ x ^0, x ^1, x ^2, x ^3 \} =\{ t, x, y, z \}$,
$f$ is a positive function and $g _{ab}$ is a symmetric two-by-two matrix.
It is convenient to introduce the null co-ordinates $( {\x} , {\y}) =
( \half ( z + t ), \half ( z - t ) )$, since the form of the metric 
{\dvakv} is preserved by the conformal co-ordinate transformations
$ ({\x} , {\y} ) \to ( \tilde\xi(\xi),\tilde\eta(\eta) )$.
The positivity of
the function $f$ is preserved if 
$\pde {\tilde\xi} {\x} \, \pde {\tilde\eta} {\y} > 0$.

The complete set of vacuum Einstein equations for the metric
{\dvakv} decomposes into two groups of equations [\BZ]. The first group
determines the matrix $g _{ab}$ and can be written as a single
matrix equation, called the Ernst equation
$$
\eqalignno{ \pde { \bigl ( \s \, \pde g {\x} \, {g ^{-1}} \bigr ) } {\y}
+ \pde { \bigl ( \s \, \pde g {\y} \, {g ^{-1}} \bigr ) } {\x} &= 0
\, ,  &\eqnalign\sigmamodel \cr}
$$
where ${\s}^2 = \det g$.
The second group of equations determines the function $f( \x , \y )$
in terms of a given solution of the Ernst equation
$$
\eqalignno{
\pde {(\ln f)} {\x} &= { \partial ^2 _{\x} {(\ln {\s})}\over{ \partial _{\x}
{(\ln {\s})} }} + { 1\over{ 4 {\s} \, {\s} _{\x} }} \tr A ^2 
\, , & \eqnalign\fa \cr
\pde {(\ln f)} {\y} &= { \partial ^2 _{\y} {(\ln {\s})}\over{ \partial _{\y}
{(\ln {\s})} }} + { 1\over{ 4 {\s} \, {\s} _{\y} }} \tr B ^2 
\, , & \eqnalign\fb \cr}
$$
where ${\s}_{\x} = \pde {\s} {\x}$, ${\s}_{\y} = \pde {\s} {\y}$ and
the matrices $A$ and $B$ are defined by
$$
\eqalignno{ A = - {\s} \ \pde g {\x} \ g ^{-1},
\quad & \quad B = {\s} \ \pde g {\y} \ g ^{-1} \, . & \eqnalign\ab \cr}
$$
Thus the dynamics of the system is determined by 
the Ernst equation {\sigmamodel}. An important consequence of the Ernst 
equation is that ${\s}_{\x \y} = 0$ so that
$\s = c(\x) + d (\y)$.
By using the conformal transformations, one can bring the functions $c({\x})$
and $d({\y})$ to a prescribed form. 

The crucial step in the inverse scattering method
is to define the linearized system whose integrability conditions
are the equations of interest, in our case the equation {\sigmamodel}. 
Following the ref. [\BZ], we define two differential operators
$$
\eqalignno{D_1 &= \pde {\z} {\x} - {{2{\s} _{\x} \, {\l}}\over{ {\l} - {\s} }} 
\, \pde {\z} {\l} \quad,\quad
D _2 = \pde {\z} {\y} + {{2{\s} _{\y} \, {\l}}\over{ {\l} + {\s} }} 
\, \pde {\z} {\l}, & \eqnalign\db \cr}
$$
where ${\l}$ is a complex parameter independent of the co-ordinates
$\{ {\x} , {\y} \}$. The
differential operators $D _1$ and $D _2$ commute since ${\s}$
satisfies the wave equation, and hence one can
consider the following linear system
$$
\eqalignno{ D_1 {\f} &= {A\over{ {\l} - {\s} }}\, {\f}
\quad,\quad
	    D_2 {\f} = {B\over{ {\l} + {\s} }}\, {\f}		
\quad ,  & \eqnalign\lsb \cr}
$$
where ${\f} ( {\l} , {\x} , {\y} )$ is a complex matrix function.
The integrability
condition for the system {\lsb} is given by the Ernst equation 
{\sigmamodel}. 
Furthermore, a solution  
${\f} ( {\l} , {\x} , {\y} )$ yields a matrix $g ( \x , \y )$ 
that satisfies the Ernst equation
{\sigmamodel}. Namely, the matrix $g ( \x , \y )$ is given by 
$$
\eqalignno { g ( \x , \y ) &= {\f} ( {\l} , {\x} , {\y} ){\big |} _{{\l}= 0} 
\ .  & \eqnalign\gpsi \cr}          
$$
In order to take into account that $g ( \x , \y )$ is real and symmetric 
we have to impose two additional conditions, see [\BZ]. Also, it is easy 
to see that the equations {\lsb} for ${\l}= 0$, imply 
equations {\ab}.

Bianchi spacetimes (see [\JAN] for a review and references) have finitely many
degrees of freedom, and only the Bianchi types I, II, 
VI$_0$ and VII$_0$ admit two commuting spacelike Killing vectors.
The metric for these Bianchi spacetimes has the form {\dvakv}, and
this can be shown by considering the general Bianchi spacetime metric 
$$
\eqalignno { ds^2 &= - dT^2 + g_{ij}(T,x^k) \ dx^i dx^j \ , 
& \eqnalign\bma \cr}
$$
where $g_{ij} = g_{IJ}(T) \  {{\vf}^I}_i (x^k) \ {{\vf}^J}_j (x^k)$ 
and $\vf^I (x^k)$ are
the one-forms associated with the spatial manifold. These one-forms satisfy 
the Maurer-Cartan equations
$$d\vf^I + \half {C^I}_{JK} \vf^J \wedge \vf^K = 0 \quad, $$
where the structure constants ${C^I}_{JK}$ correspond to the Lie algebra of the
symmetry group of the Bianchi model.
For the relevant models the structure constants satisfy
$$ 
\eqalignno {
{C^I}_{JK} & = \e_{JKL} S^{LI} \quad,& \eqnalign\csc \cr}
$$
where $\e_{JKL}$ is a totally antisymmetric tensor density
and $S$ is a symmetric matrix.
In this case the one forms ${\vf}^I$ take 
the following form 
$$
\eqalignno { {\vf} ^1 &= {l ^1} _1(z) \ {\dif} x + {l ^1} _2 (z) \ {\dif} y 
\ , \quad 
             {\vf} ^2  = {l ^2} _1 (z) \ {\dif} x + {l ^2} _2 (z)\ {\dif} y 
\ , \quad
             {\vf} ^3  = {\dif} z  \ .  & \eqnalign\baforms \cr}
$$
An important consequence of the  Maurer-Cartan equations for the one forms 
${\vf} ^I$
is that the matrix $l= ||{l^a}_b ||$ satisfies the following linear 
differential equation
$$
\eqalignno { {dl\over dz} &= C^T \e l \quad, &\eqnalign\bsa \cr}
$$
where the matrix $C$ is the upper two-by-two block on the 
principal
diagonal of the matrix $S^{IJ}$ and $\e$ is 
the antisymmetric matrix with $\e_{12} = 1$.

After a time redefinition $t = t (T)$, the metric {\bma} can be written 
in the form {\dvakv} 
$$
\eqalignno { ds^2 &= f(t) \ (-dt^2 + dz^2 ) + g_{ab}(t,z) \ dx^a dx^b \quad, 
&\eqnalign\csm \cr}
$$
where
$$
\eqalignno { g ( t , z ) &= l^T (z) \hat\g (t) l (z) \quad, &\eqnalign\hr \cr}
$$
and $\hat\g$ is a two-by-two symmetric matrix. 
Notice that now $ \s^2 = ( \det l )^2 \det \hat\g $, and 
since $\det l = 1$ we get
$$
\eqalignno { \s^2 ( t )&= \det \hat\g ( t) \ . &\eqnalign\al \cr}
$$
In addition, $\s$ has to satisfy the wave equation, so that
$ \ddot {\s } ( t )= 0$, and
hence $\s$ can only be a linear function of time.  

The linearized system 
{\lsb} can be simplified for the models described by the metric {\csm}. 
The first step is to define a two-by-two matrix function $\varphi$ by
$$
\eqalignno { \psi (t,z,\l) &= l^T (z) \, \varphi (t,z,\l) \ l (z) \ , 
& \eqnalign\hphi \cr}
$$
and a constant two-by-two matrix $R = \e \ C$.
The second step is to substitute {\hr} into {\ab} and use the definition
of the coordinates $\x$ and $\y$. Then the results of these calculations,
together 
with the definition {\hphi}, can be used to simplify the equations 
{\lsb}.
The crucial step in which a simplification occurs is to perform a 
conformal coordinate
transformation $\{ t , z , \l \} \to \{ t , w , \l \}$, where $w$ is given by
$w = {\half} \bigl ( {\s^2\over \l} + 2 \b + \l \bigr )$.
The linear system after this coordinate 
transformation 
involves only derivatives in $t$ and $\l$ since all the terms involving 
derivatives in $w$ are canceled. Finally, it is useful to make some simple 
linear combinations of the two equations and to use the fact that 
$\s$ is a linear function of time. In this way one obtains a new linear system
$$
\eqalignno { \pde \varphi t  &= {t\over \l} \ \bigl ( \hat\g R ^T 
{\hat\g}^{-1} \varphi
- \varphi R ^T	\bigr ) \quad , \cr 
\pde \varphi \l &=  {1\over 2} \ \bigl ( - R \varphi - \varphi R ^T
+ {t\over \l} {\dot \g} {\hat\g}^{-1} \varphi + {t^2\over \l ^2} \varphi R ^T
- {t^2\over \l ^2} \hat\g R ^T {\hat\g}^{-1} \varphi \bigr )
\quad , &\eqnalign\lsh \cr}
$$
where we have set $\s =t$.

Although the matrix function $\varphi ( t , \l , w)$ depends 
on all 
three variables, the right-hand side of the system {\lsh} does not have any 
$w$ dependence.  
The integrability condition for the system {\lsh} is 
$$
\eqalignno { {1\over t}{d\over dt}\bigl( t \dot{\hat\g} {\hat\g}^{-1}\bigr)  
&= R\hat\g R^T {\hat\g}^{-1}
- \hat\g R^T {\hat\g}^{-1} R \quad . &\eqnalign\zcb \cr}
$$

Equivalently, one can derive the equation {\zcb} by a direct substitution of 
the formula {\hr} into  equation {\sigmamodel}. Thus the dynamics of 
these Bianchi models is determined by the equation {\zcb}. 

{\bf\chapter{Zero-curvature representation for Painlev\'e III}}

The linear system {\lsh} and the corresponding nonlinear equation 
{\zcb} were derived for special matrices $R$, which correspond to $ C$
matrices of the relevant Bianchi models. Note that the $C$ matrix is symmetric,
and the only relevant information about the Bianchi model is contained in its
signature, so that there are four distinct possibilities:

(1) $C = diag(1, 1)$ for Bianchi VII$_0$,

(2) $C = diag(1, -1)$ for Bianchi VI$_0$,

(3) $C = diag(1, 0)$ for Bianchi II,

(4) $C = diag(0, 0)$ for Bianchi I.

\noindent However, if we consider the linear system {\lsh} independently of 
Bianchi models, then we can take $C$ to be an arbitrary symmetric
two-by-two matrix. In this case we have
$$
\eqalignno { C &=  \pmatrix { c & d \cr
			      d & k \cr}
\ ,  &\eqnalign\chat \cr}
$$  
with $R = \e C$, $\hat\g = diag (a , b)$ and $ab = t^2$. 
Then the consistency condition {\zcb} gives
$$
\eqalignno {   
t^{-1}{d\over dt}(t a^{-1} \dot a ) &= k^2 a^{-1}b - c^2 ab^{-1}  \cr
		0	&= d ( k + c ab^{-1})
\ .  &\eqnalign\zcc \cr}
$$  
The second equation implies either $d=0$ or $a^2 = -(k /c)t^2$. The second
possibility gives a linear in time solution, and the first possibility is more
interesting. $d=0$ case gives 
$$
\eqalignno {   
t^{-1}{d\over dt}(t a^{-1} \dot a ) &= k^2 a^{-2}t^2 - c^2 a^2 t^{-2}  
\quad .  &\eqnalign\pgen \cr}
$$   
By making a change of variables $u=a^2 /t^2$, $\t = t^2 /4$, the equation
{\pgen} takes the standard Painlev\'e III form {\spiii}
$$
\eqalignno { {d ^2 u\over d \t ^2} &= {1\over u} \Bigl ( {d u\over d \t} 
\Bigr ) ^2 - {1\over \t} \Bigl ( {d u\over d \t} \Bigr ) + {2\over \t}
 \Bigl ( -c^2 u^2 + k^2 \Bigr ) \ , &\eqnalign\piiic \cr}
$$
so that $\a = -2c^2$, $\b = 2k^2$ and $\g =\d =0$.

Let us now consider the Bianchi models. In the case of Bianchi VII$_0$ model
the spatial 
hyper-surface is a three torus $T^3$. The matrix $C$ is $diag (1,1)$ and 
the matrix $l$ is  given by
$$
\eqalignno {   l(z) &=  \pmatrix { \cos z & \sin z \cr
			      - \sin z & \cos z \cr}
\quad .  &\eqnalign\matrixlbs \cr}
$$
The matrix $R$ is given by 
$$
\eqalignno {   R &=  \pmatrix { 0 & 1 \cr
			      - 1 & 0 \cr}
\ ,  &\eqnalign\matrixrbs \cr}
$$
so that the local dynamics is given by {\pgen} for $c=k=1$. By making a 
change of variables $u = e^q$, this PIII equation takes a more symmetric form
$$
\eqalignno {
{d\over d\t} \left(\t {dq\over d\t} \right) &=- 4\sinh q  \quad.
 &\eqnalign\symp \cr}
$$

Bianchi VI$_0$ model corresponds to $C=diag(1,-1)$, so that the equation 
{\bsa} gives
$$
\eqalignno {l(z) &=  \pmatrix { \cosh z & \sinh z \cr
			      \sinh z & \cosh z \cr}
\quad .  &\eqnalign\lbsx \cr}
$$
This model has a non-compact spatial manifold, which is locally compact, 
and the local dynamics is the same as in the
Bianchi VII$_0$ case, because $c^2 = k^2 =1$. 

In the Bianchi II case $C=diag(1,0)$, and the dynamics is given by {\pgen}
with $c=1$ and $k=0$. The $l$ matrix is given by
$$
\eqalignno {l(z) &=  \pmatrix { 1 &  z \cr
			      0 & 1 \cr}
\quad ,  &\eqnalign\ltwo \cr}
$$
and this model allows compact spatial sections [\K]. 
In the Bianchi I case $C=0$, so that $c=k=0$ and $l = diag (1,1)$.

{\bf \chapter {Conclusions}}

We have shown that a class of PIII equations ($\a>0,\b<0,\g=\d=0$) arises as 
a zero-curvature condition
in the Belinskii-Zakharov inverse scattering method applied to Bianchi 
spacetime metrics. For the particular 
values of the parameters these PIII equations become the dynamical equations 
for Bianchi I,
II, VI$_0$ and VII$_0$ models. Note that the
linear system {\lsh} can be transformed into the standard form 
$$ 
\eqalignno {   
{\partial\Psi \over \partial\l} &= {\hat A}\Psi \quad,\quad 
{\partial\Psi \over \partial t} = {\hat B}\Psi 
&\eqnalign\blsn \cr}
$$
where $\hat A$ and $\hat B$ are four-by-four matrices given by
$$
\eqalignno {   
\hat A &= {t\over \l} \ \bigl ( \g R^T \g^{-1} \otimes I_2
+ I_2 \otimes R^T \bigr ) \cr  
\hat B &= {1\over 2} \ \bigl ( - R  
+ {t\over \l} {\dot \g} \g ^{-1} 
- {t^2\over \l ^2} \g R ^T \g ^{-1} \bigr )\otimes I_2 + 
\half \left({t^2\over \l^2}-1 \right)I_2 \otimes R \quad,
&\eqnalign\zcn \cr}
$$
where $I_2$ is the identity matrix and $\Psi$ is a column formed from the 
columns of the matrix $\varphi$.
 
This linear system is different from the 
linear system which is
used for the study of Painlev\'e III equation within the isomonodromic 
deformation (IMD) method 
[\FN,\JMU,\I]. Although it is not obvious what are the advantages of the new
linear system in comparison with the IMD linear system, there are some 
interesting features of the new system which can be investigated.

In the Belinskii-Zakharov inverse scattering approach, it is natural to 
consider path-ordered exponentials of matrices (holonomies) associated to 
the linear system {\lsb} 
in order to find the integrals of motion [\MS],[\KS]. 
In the special case of Bianchi metrics, the holonomy construction
would simplify, and one could try to see what kind of expressions one would
obtain for the PIII equation. 

Note that the new linear system can be
interpreted as a Lax pair for a  dynamical system with a time-dependent
Hamiltonian
$$
\eqalignno {   
H &= {1\over 2\t} p^2 + 2 (k^2 e^{-q} + c^2 e^q )\quad.
&\eqnalign\tdh \cr}
$$
This is analogous and complementary to the 
results of Harnad and Routhier [\HR], where a Lax pair for PIII equation with 
$\a\b\g\d \ne 0$ was constructed. It is interesting that in that case the
Lax pair contains $\int d\t u$, which does not happen in our case of PIII
equation. 

In the context of Bianchi models, it is more natural to work with
a dynamically equivalent Hamiltonian to {\tdh}  
$$
\eqalignno {   
\tilde H &= \half {\tilde p}^2 + 2 e^{\tilde t} (k^2 e^{-q} + c^2 e^q )\quad,
&\eqnalign\tdhn \cr}
$$
where $\t = e^{\tilde t}$. One can now examine the physical properties of the 
solutions, like small and large time asymptotic, as well as the singularities, 
since these properties of the Painlev\'e III solutions
have been thoroughly studied [\I]. 


\noindent{\bf Acknowledgements}

N.M. was partially supported by the grant
PRAXIS/2/2.1/FIS/286/94 and A.M. was supported by the grant 
PRAXIS XXI/BCC/18981/98 from the Portugese Foundation for
Science and Technology.

\refout

\bye